\documentclass[english]{IEEEtran}
\usepackage[T1]{fontenc}
\usepackage[latin9]{inputenc}
\usepackage{geometry}
\usepackage{textcomp}
\usepackage{amsmath}
\usepackage{graphicx}

\makeatletter
\@ifundefined{showcaptionsetup}{}{%
 \PassOptionsToPackage{caption=false}{subfig}}
\usepackage{subfig}
\makeatother

\usepackage{babel}
\begin{document}
\title{Observation of a Pinched-Loop in a Current-Excited Inductive Circuit}
\author{Ahmed S. Elwakil\textsuperscript{{*}§}\thanks{{*} Department of Electrical and Computer Engineering, University
of Sharjah, PO Box 27272, Sharjah, United Arab Emirates and Department
of Electrical and Software Engineering, University of Calgary, Calgary,
Alberta, T2N 1N4 }\thanks{§ Department of Electrical and Software Engineering, University of
Calgary, Calgary, Alberta, T2N 1N4}, Costas Psychalinos\textsuperscript{§§ }\thanks{§§ Department of Physics, Electronics Laboratory, University of Patras,
Rio Patras, GR-26504, Greece}, Brent J. Maundy\textsuperscript{§}and Anis Allagui\textsuperscript{{*}{*}}\thanks{{*}{*} Department of Sustainable and Renewable Energy Engineering,
University of Sharjah, PO Box 27272, Sharjah, United Arab Emirates }}
\maketitle
\begin{abstract}
In this work, we show that a pinched-loop can be observed in the voltage-current
plane when a series R-L circuit is current excited. Specifically,
the resistance (R) in this circuit is variable and is voltage-controlled
by the voltage developed across the inductor due to the exciting current.
In this context, we confirm our previous results that the pinched-loop
is not a characteristic of memrsitors or memrsitive systems and that
it can be observed in many other nonlinear systems. Numerical simulations,
circuit simulations and experimental results validate the theory.
\end{abstract}

\section{Introduction}

Memristors have been studied as new electronic devices with promising
analog/digital and neuromorphic applications \cite{app1,app2,bio}.
Their characteristic behavior is a pinched-loop in the current-voltage
plane \cite{elwakil2013simple}. This behavior however has been shown
to exist in other nonlinear devices such as a non-linear inductor
or a non-linear capacitor \cite{fouda2015pinched}, and that its appearance
is linked to satisfying the necessary conditions of the theory of
Lissajous figures \cite{maundy2019correlation}. The pinched-loop
is a Lissajous figure of order two; i.e. with one point of intersection
(also known as the pinch point). To generate such a figure from two
signals, one of them must contain a harmonic component with twice
the frequency of the other signal. In addition, the phase shift between
this harmonic and the fundamental frequency component must not exceed
$\pm\pi/2$ \cite{maundy2019correlation}. Without using a nonlinear
device, the frequency-doubling mechanism mandated by the theory of
Lissajous figures, cannot be generated from an applied input signal
(voltage or current) with a fixed frequency. We have recently verified
this fact using commercially available memristor devices in \cite{MAJZOUB}
where we have shown that the measured memristor current signal does
indeed contain a strong harmonic component at twice the frequency
of the applied voltage signal. This harmonic is the only one needed
to generate the pinched-loop and therefore all other higher-order
harmonics can be filtered out from the current signal \cite{MAJZOUB}.

In this work, we show that a current excited series R-L network can
also generate a pinched-loop if the resistor is nonlinear and voltage-controlled
by the voltage developed across the inductor resulting from the current
excitation. This forms a kind of self-feedback state-control which
in-directly implements a multiplication operation. This multiplication
operation generates a harmonic component (in the voltage developed
across the R-L circuit) with twice the frequency of the applied input
current signal. The proposed concept is validated using a simple circuit
implementation that relies on a junction field effect transistor (JFET)
operating as a voltage-controlled resistor. When compared to other
similar concepts in the literature \cite{alharbi2017electrical},
the robustness and simplicity of this work become apparent. An experimental
verification using a Field Programmable Analog Array (FPAA) of the
proposed concept is also provided.

\section{Theory}

Consider the circuit shown in Fig. 1 composed of an inductor in series
with a resistor which has a fixed part ($R_{f}$) and a variable part
($R_{v}$). The network is excited by a current source ($i_{in}$)
and the developed voltage on the network ($v_{in}$) is measured.
The variable resistance $R_{v}$ is voltage-controlled by the inverted
voltage developed on the inductor; i.e. $-L\frac{di_{in}}{dt}$ or
a scaled (by factor $m$) version of this voltage. Therefore, the
total resistance of the network is given by 
\begin{figure}
\begin{centering}
\includegraphics[scale=0.25]{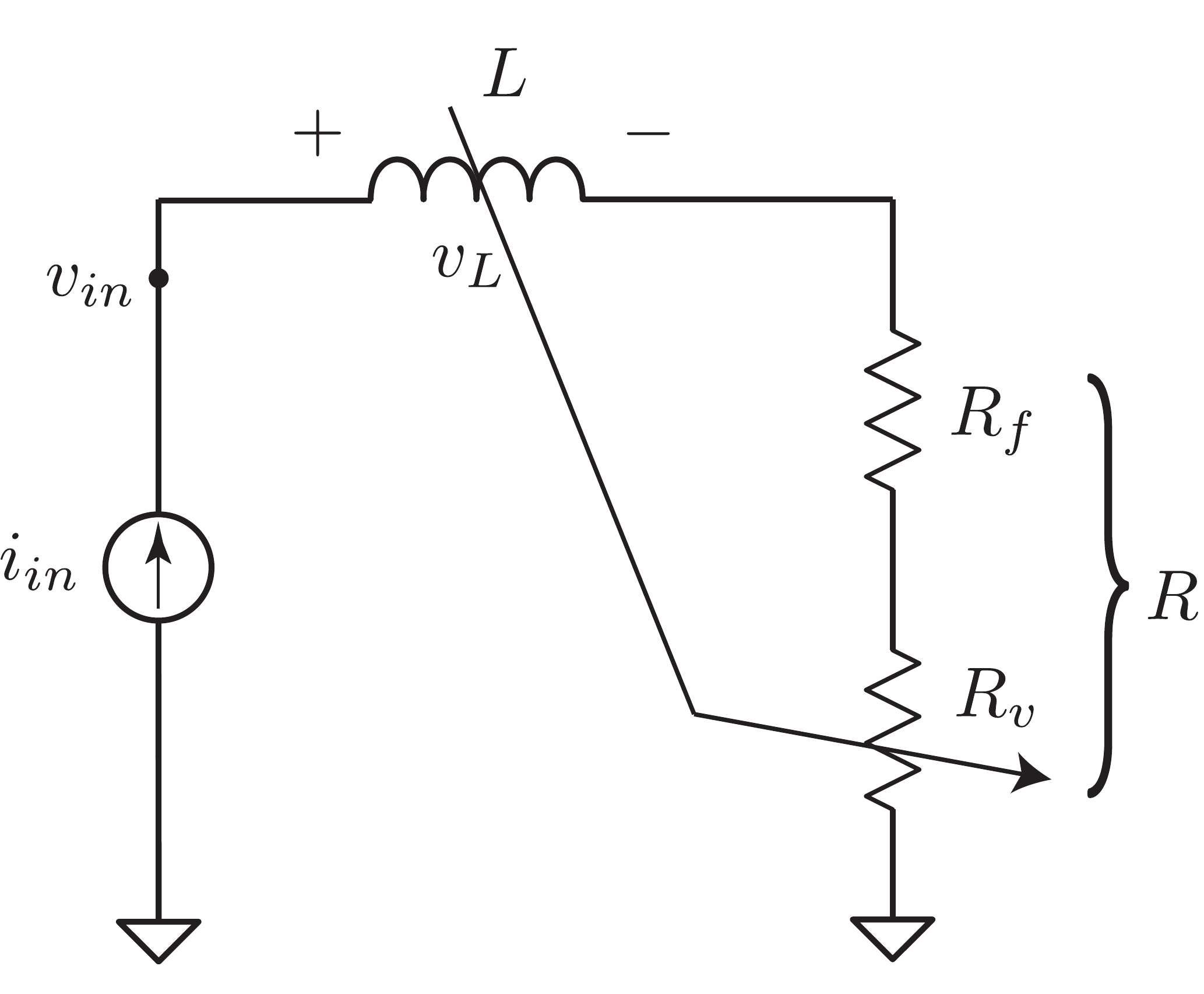}
\par\end{centering}
\caption{Current excited R-L circuit with voltage-controlled resistor $R_{v}$
controlled by the flux-induced voltage $v_{L}$ developed on the inductor}
\end{figure}
\begin{equation}
R=R_{f}-m\cdot R_{v}\cdot L\frac{di_{in}}{dt}\cdot\frac{1}{v_{ref}}
\end{equation}
where $v_{ref}$ is an arbitrary reference voltage to ensure proper
dimensions. The input voltage measured across the network is therefore
given by
\begin{equation}
v_{in}=i_{in}R_{f}+L\frac{di_{in}}{dt}\left(1-m\frac{R_{v}i_{in}}{v_{ref}}\right)
\end{equation}
The above equation can be transformed into dimensionless form after
defining $y=v_{in}/v_{ref}$, $x=R_{v}i_{in}/v_{ref}$, $a=R_{f}/R_{v}$,
$\tau=L/R_{v}$ and using the normalized time $t_{n}=t/\tau$. We
thus obtain the following equation
\begin{equation}
y=ax+\dot{x}(1-mx)\label{eq:sys}
\end{equation}
The second term in the above equation represents a clear multiplication
operation without need for a physical multiplier. This operation is
the key behind the frequency-doubling mechanism and hence generating
the pinched-loop as demonstrated below. 
\begin{figure}
\begin{centering}
(a)\includegraphics[scale=0.3]{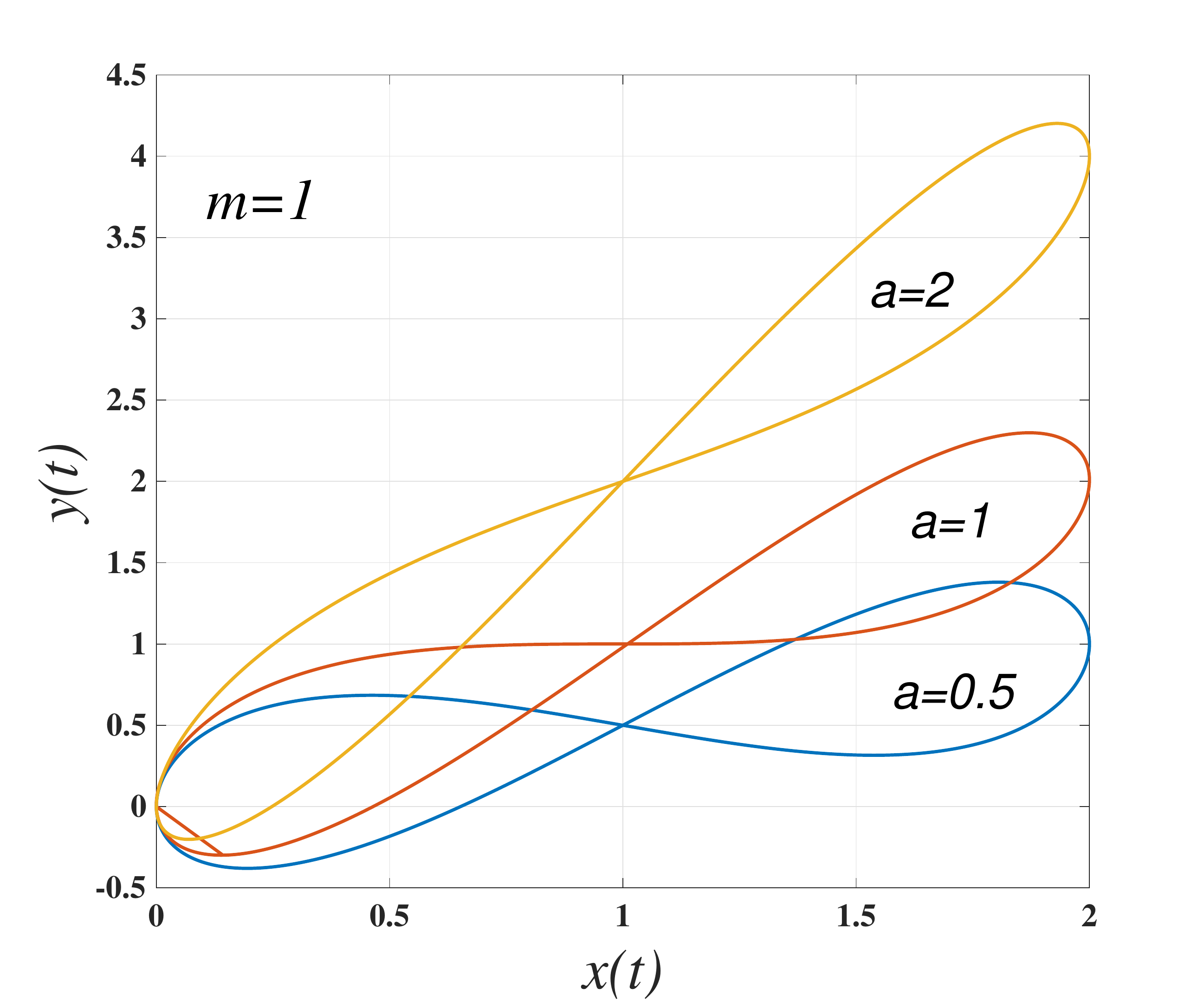}
\par\end{centering}
\begin{centering}
(b)\includegraphics[scale=0.32]{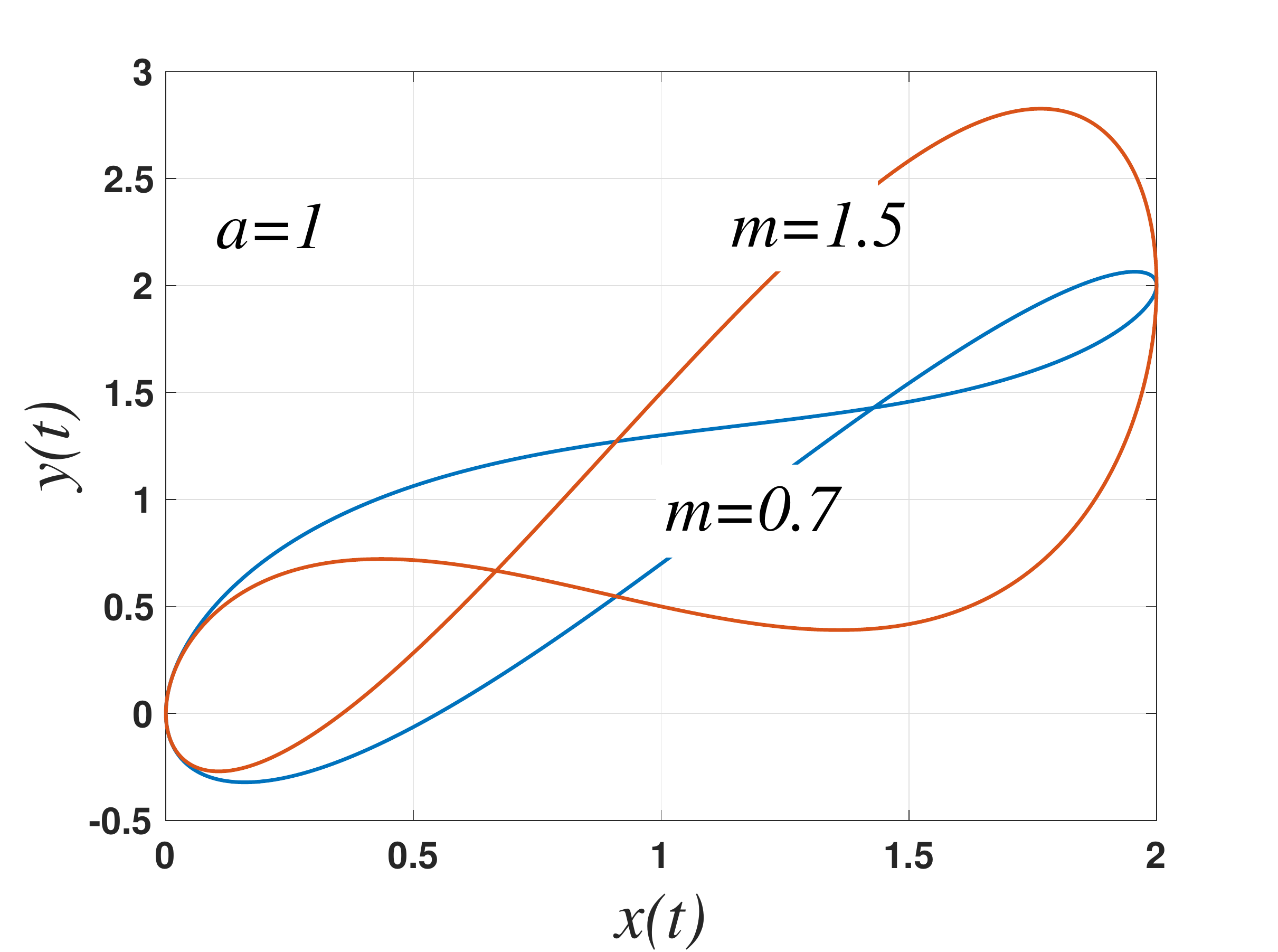}
\par\end{centering}
\begin{centering}
(c) \includegraphics[scale=0.5]{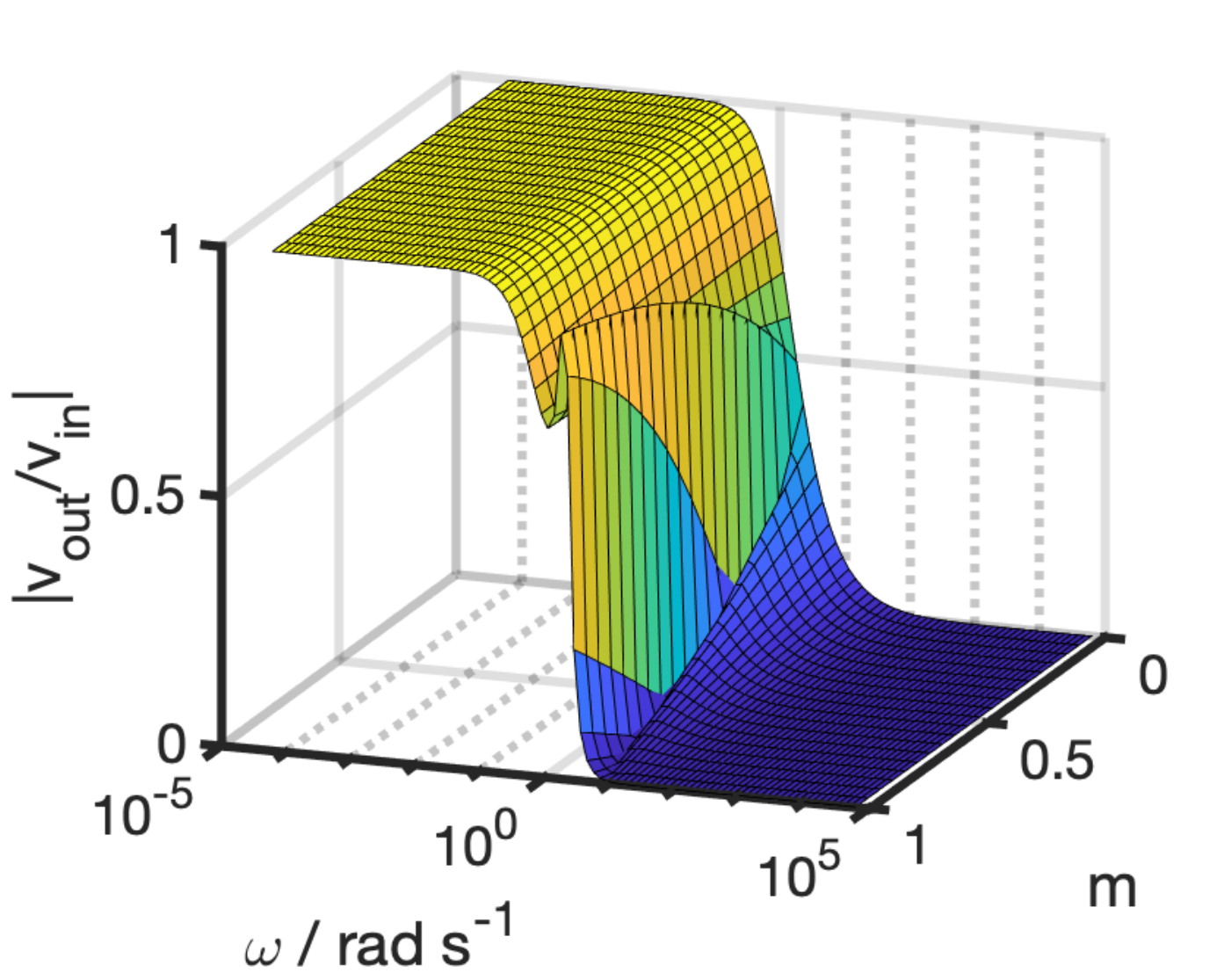}
\par\end{centering}
\caption{Pinched loops from equation (\ref{eq:sys}) for (a) $m=1$ and three
different values of $a$, (b) $a=1$ and two different values of $m$.
Magnitude response of (\ref{eq:nonlinear filter}) at $a=1$ versus
$m$ is plotted in (c)}
\end{figure}

Numerical simulation results of equation (\ref{eq:sys}) are shown
in Fig. 2(a) for $m=1$ and three different values of $a$. Here,
$\dot{x}=\sin(\omega t_{n})$ with $\omega=1$ for simplicity. The
pinched-loop behavior can clearly be obtained from equation (\ref{eq:sys})
and the pinch point is located at $(x_{p},y_{p})=(\frac{1}{m},\frac{a}{m})$.
The loop is always symmetrical in this case whereas in Fig. 2(b),
non-symmetrical loops are observed for $a=1$ and $m\neq1$. Specifically,
two results are shown for $m<1$ and $m>1$. The generation of the
pinched-loop in (\ref{eq:sys}) can be attributed to the frequency
doubling mechanism resulting from the $\dot{x}\times x$ term. For
$a=m=\omega=1$, the system is described by the two parametric equations
\begin{equation}
\begin{array}{c}
x(t_{n})=1-\cos(t_{n}),\\
y(t_{n})=1-\cos(t_{n})+\frac{1}{2}\sin(2t_{n})
\end{array}
\end{equation}
and at the pinch-point $(x_{p},y_{p})=(1,1)$, the phase shift between
the two signals is $\pi/2$ satisfying the Lissajous condition for
creating a pinched-loop. It can also be shown in this case that 
\begin{equation}
y=x+(1-x)\sqrt{2x-x^{2}}\label{eq:m1}
\end{equation}
and hence
\begin{equation}
\frac{dy}{dx}=1+\frac{2x^{2}-4x+1}{\sqrt{2x-x^{2}}}
\end{equation}
Therefore, the infliction point at $\frac{dy}{dx}=0$ is located at
$x=1$ with the corresponding value of $y=1$, as expected. Note however
that the second term in (\ref{eq:m1}) cancels out at $x=0$ and at
$x=2$. Therefore, there are two more points at which $y=x$ which
are $(0,0)$ and $(2,2)$. These two points are clearly visible in
Fig. 2(a) for $a=1$. In fact, for any value of $m$ and $a$, the
pinched loop always passes through the two points $(0,0)$ and $(2,2a)$
as clear from Fig. 2(a,b).

It is worth mentioning that taking the voltage across the resistor
$R$ in Fig. 1 as an output ($v_{out})$ and considering the magnitude
response of $v_{out}(s)/v_{in}(s)$ ; where $s=j\omega$ shows a classical
low-pass filter response with a cutoff frequency $\omega_{c}=R_{f}/L$
when $m=0$. Now with $m\neq0$, it is seen that the normalized output
voltage is equal to $y(t)-\dot{x}(t)=x(t)\left(a-m\dot{x}(t)\right)$.
Therefore assuming that $x(t_{n})=\sin(t_{n})$, it can be shown that
\begin{equation}
\frac{v_{out}(s)}{v_{in}(s)}=\frac{\mathcal{L}\left(y(t_{n})-\dot{x}(t_{n})\right)}{\mathcal{L}y(t_{n})}=\frac{\frac{a}{s^{2}+1}-\frac{m}{s^{2}+4}}{\frac{s+a}{s^{2}+1}-\frac{m}{s^{2}+4}}=\frac{a}{s+a}|_{m=0}\label{eq:nonlinear filter}
\end{equation}
since $a=R_{f}/R_{v}$ and time is normalized with respect to $\tau=L/R_{v}$,
it can be seen that the low-pass filter transfer function at $m=0$
is $\frac{1}{j(\frac{\omega}{\omega_{c}})+1}$ as expected. Figure
2(c) shows the magnitude response of (\ref{eq:nonlinear filter})
for different values of $m$ and for $a=1$.

\section{Circuit Verification}

\subsection{with discrete components}

It is possible to realize the circuit in Fig. 1 using a JFET as a
voltage-controlled resistor as shown in Fig. 3. Here, a fixed $135\Omega$
resistor is used as $R_{f}$ whereas the current feedback op amp (CFOA)
$U_{2}$ is used as a differential amplifier to obtain the voltage
developed across the inductor and re-inject it to the gate terminal
of the JFET device. The control voltage for the JFET resistance is
thus equal to $-(R_{3}/R_{2})v_{L}$. Meanwhile, op amp $U_{1}$ is
another CFOA used as a voltage-to-current converter in order to obtain
the excitation current $i_{in}=v_{i}/R_{i}$ through its high impedance
current output terminal. 
\begin{figure}
\begin{centering}
\includegraphics[scale=0.35]{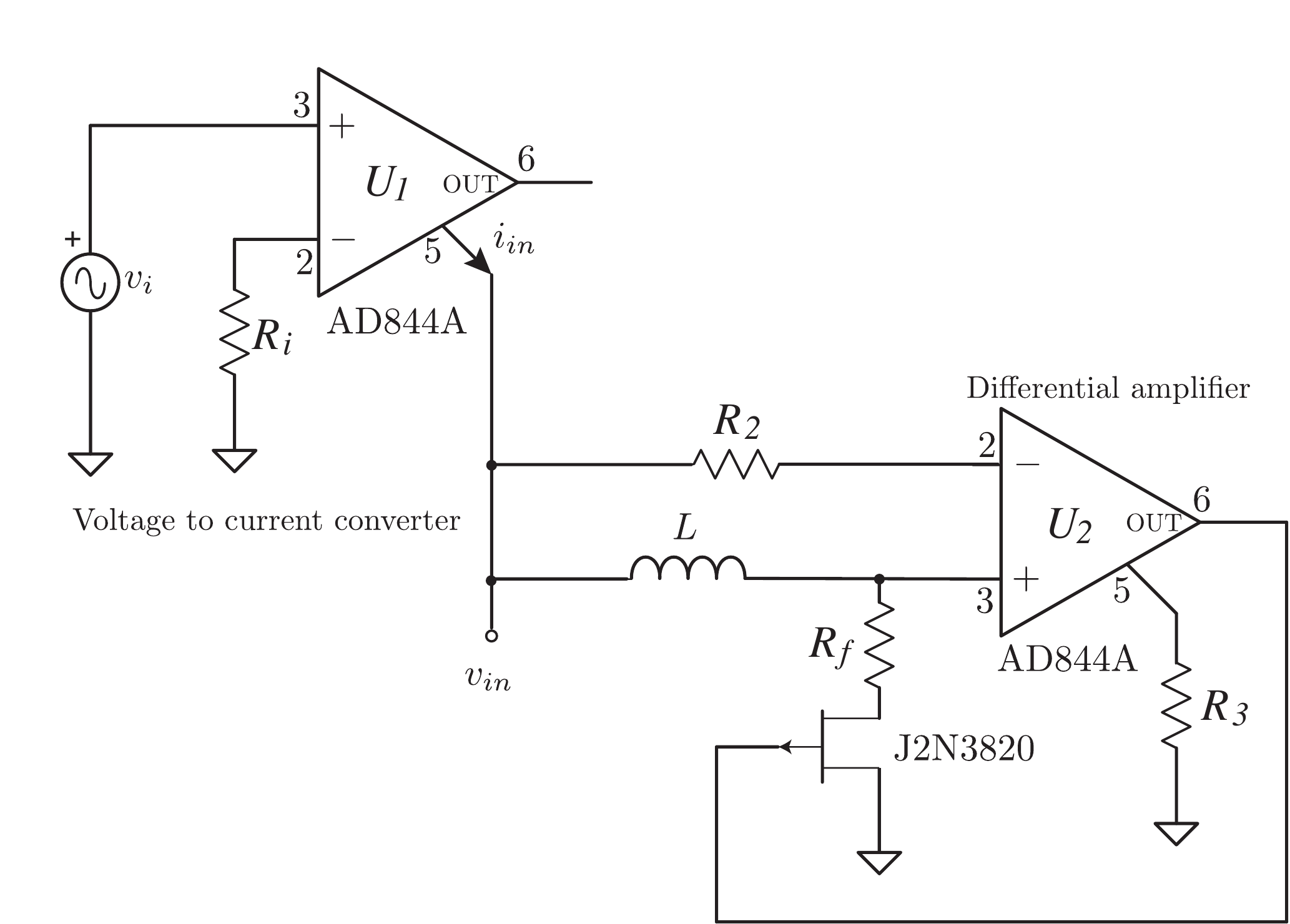}
\par\end{centering}
\caption{Schematic of the simulated circuit realizing the concept in Fig. 1}
\end{figure}
 
\begin{figure}
\begin{centering}
\includegraphics[scale=0.45]{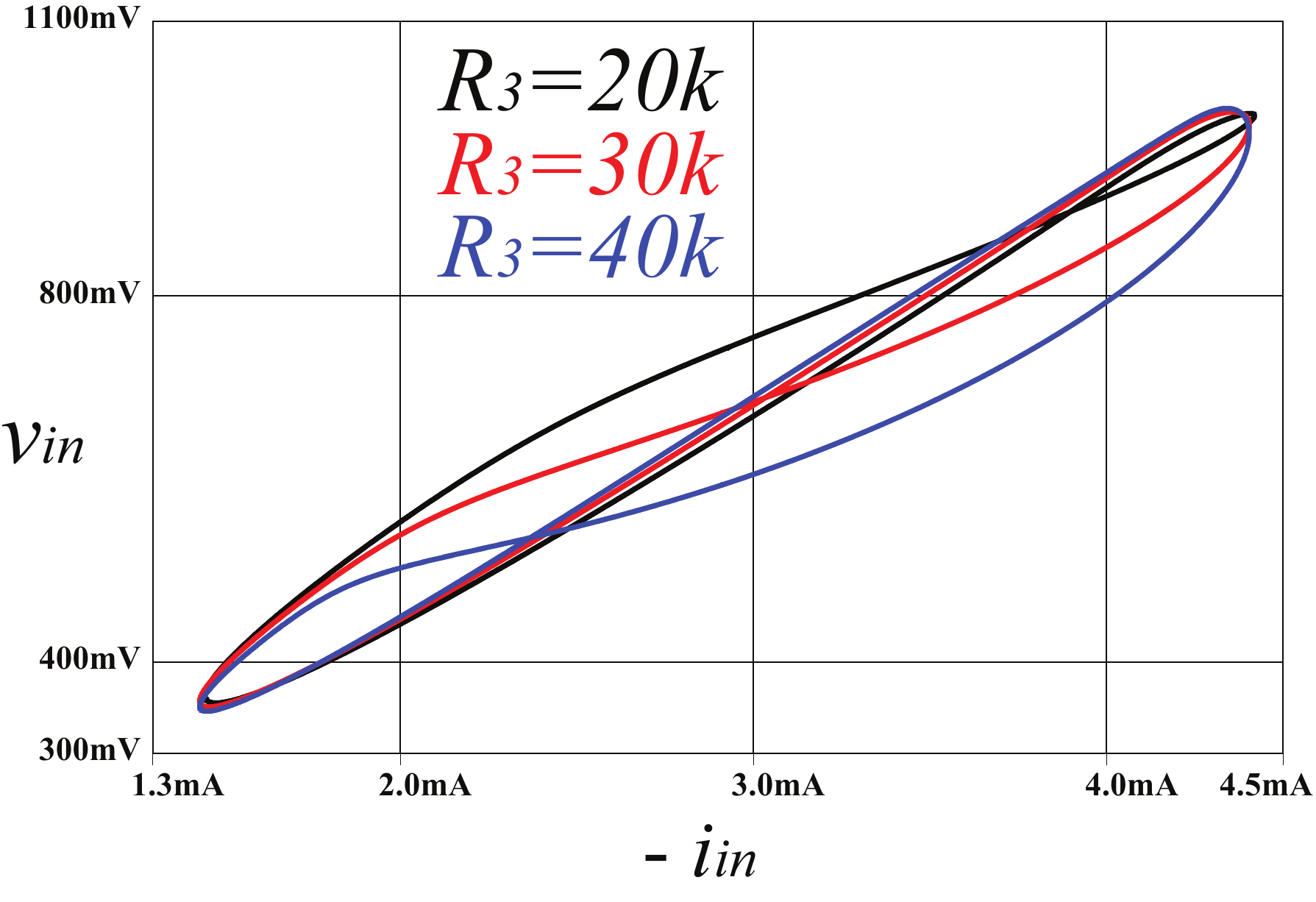}
\par\end{centering}
\caption{Spice simulated pinched-loop from the circuit in Fig. 3 for $R_{i}=1k\Omega$,
$R_{2}=10k\varOmega$ and three different values of $R_{3}$ ($R_{3}=20k\Omega,30k\varOmega,40k\Omega$)
corresponding respectively to the black, red and blue loops. Op amps
$U_{1}$ and $U_{2}$ were biased with $\pm9V$ supplies}
\end{figure}

Using an AD844 CFOA device and a J2N3820 JFET device, the Spice simulated
response of the circuit is plotted in Fig. 4 depicting $i_{in}$ versus
$v_{in}$ for three different ratios of $R_{3}/R_{2}$ namely $2,3$
and $4$. The inductor was chosen as $L=10mH$ and $v_{i}$ had an
amplitude of $1.5V$ with a DC offset voltage of $+3V$ and a frequency
of $1kHz$. The nominal resistance of the JFET was measured as $R_{v}\approx135\varOmega$
and therefore the circuit simulations correspond to $a=R_{f}/R_{v}\approx1$
in the numerical simulations. We noted that at $R_{3}=30k\Omega$,
a symmetrical loop is observed (see red loop in Fig. 4) whereas for
a lower and a higher value, the loop is not symmetrical. Therefore,
$(R_{3}/R_{2})$ plays the role of $m$ and controls the loop symmetry
when $a=1$. 

\subsection{with a Field Programmable Analog Array}

\begin{figure}
\centering{}\includegraphics[scale=0.2]{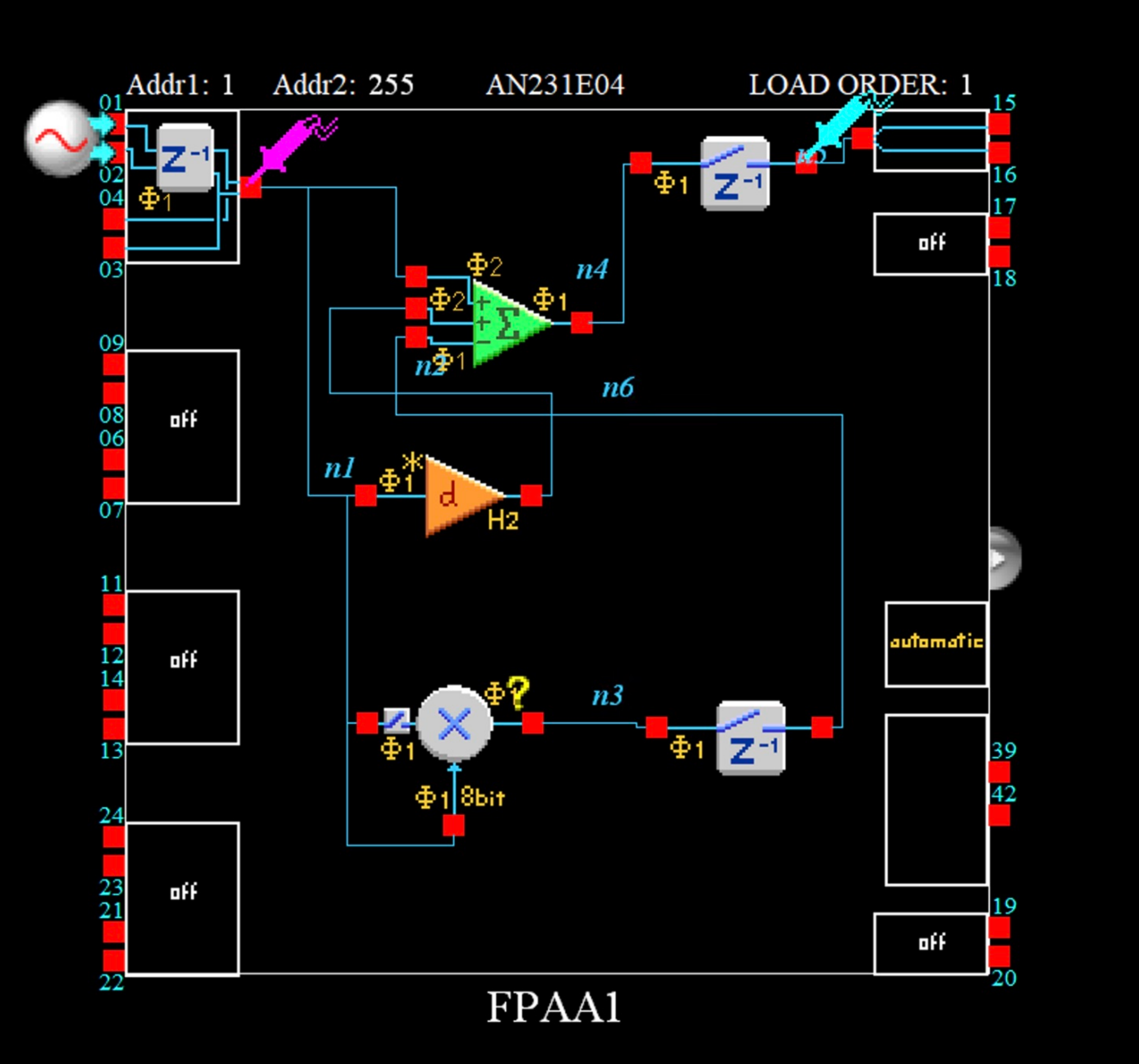}\caption{FPAA configuration for experimentally verifying equation (\ref{eq:sys})
in voltage mode input/output \label{fig:FPAA}}
\end{figure}

\begin{figure}
\noindent \begin{centering}
\subfloat[\label{fig:variable_alpha}]{\centering{}\includegraphics[scale=0.42]{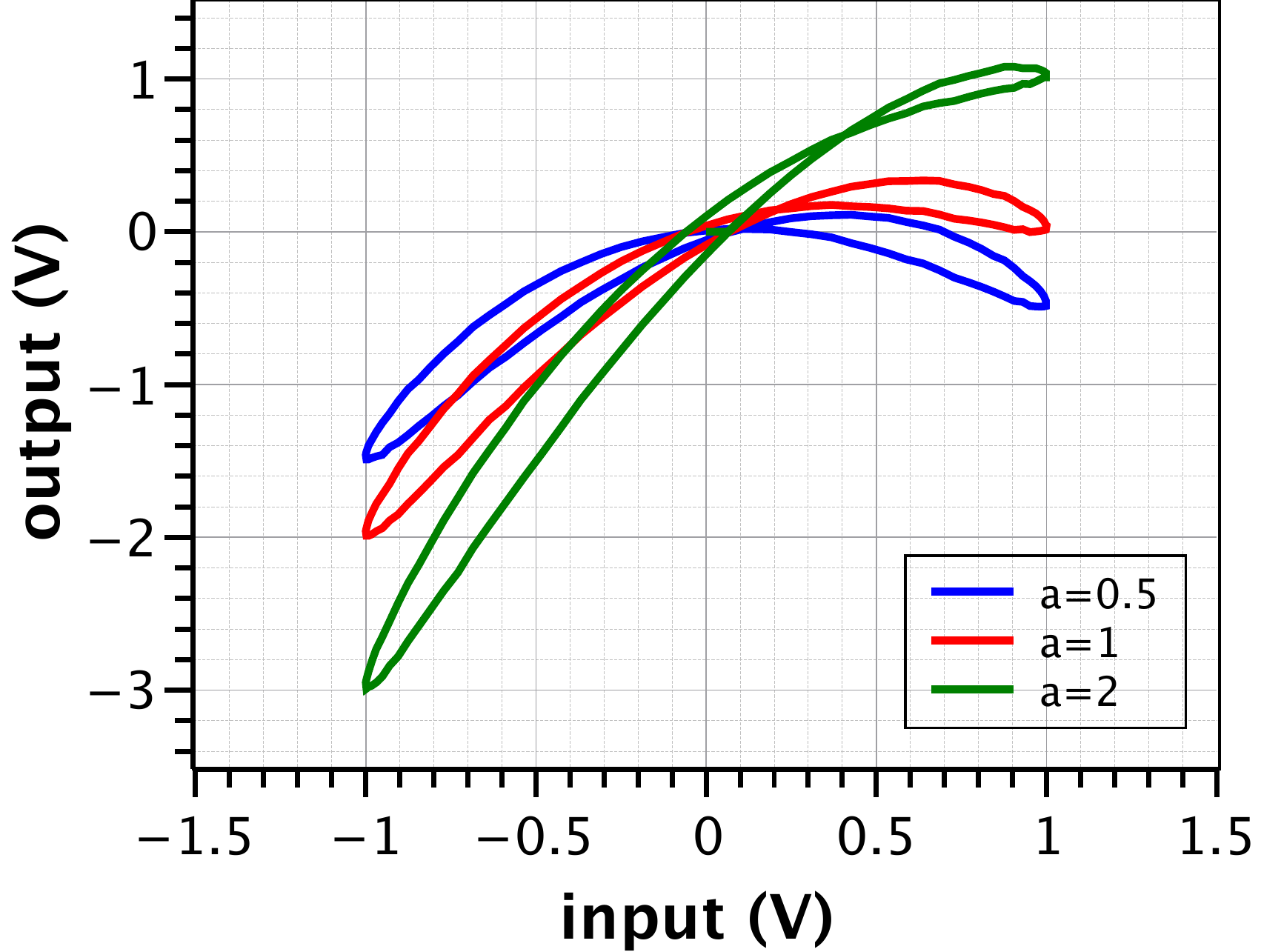}}
\par\end{centering}
\noindent \centering{}\subfloat[\label{fig:variable_mu}]{\centering{}\includegraphics[scale=0.42]{fig6a.pdf}}\caption{Experimental pinched-loops from the FPAA realization for (a) $m=1$
and $a=0.5,1,2$ and (b) $a=1$ and $m=0.7,1.5$ in equation (3)}
\end{figure}

For experimental verification, the AN231E04 FPAA from Anadigm was
employed to validate the concept described by (\ref{eq:sys}). The
board is programmed through the\textit{ AnadigmDesigner}\textsl{}\textsuperscript{\textsl{®}}\textsl{2}
EDA software, and the tested configuration is depicted in Fig. \ref{fig:FPAA},
where the \textit{Hold, SumDiff }(summation stage), \textit{Differentiator}
and \textit{Multiplier} CAMs are utilized. The clock frequency was
set to 100kHz, while the system is stimulated by a 1KHz, 1V input
signal. Considering that $m=1$ and $a=0.5,1,2$ the results of the
synthesis of the input and output waveforms are demonstrated in Fig.
(\ref{fig:variable_alpha}), while the corresponding results for $a=1$
and $m=0.7,1.5$ are provided in Fig. (\ref{fig:variable_mu}). The
skewness in the pinched-loops is attributed to the limited dynamic
range of the FPAA. 
 
\section{Discussion}

We stress that pinched-loop behavior can only exist in nonlinear circuits
and systems meaning that any fabricated solid-state device that shows
a pinched-loop must have an embedded form of state-controlled nonlinearity;
most likely a voltage- or current-controlled resistor or transconductor.
In this regards, it is important to recall that an inductor may exist
in the form of a parasitic element not intentionally inserted in series
with a voltage-controlled resistor. If the voltage developed across
this inductor leaks to the resistor (e.g. through magnetic coupling),
it may then cause the appearance of the pinched-loop behavior. Interestingly,
in the very recent work of \cite{N1}, the author show that \emph{``the
halide perovskite memristor response contains the composition of two
inductive processes}''. Therefore, the existence of inductive behavior
in nano-materials is not unusual. 

\begin{figure}
\centering{}\includegraphics{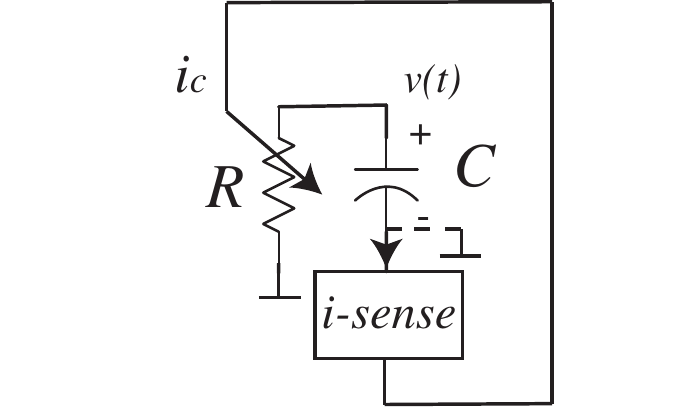}
\caption{\label{fig:The-dual-of}The dual of the R-L circuit in Fig. 1 is a
parallel R-C circuit where the resistor is controlled by the current
developed in the capacitor}
\end{figure}

Furthermore, applying the duality principle to the circuit in Fig.
1, it can be established that a voltage-excited parallel R-C circuit
where the resistor is controlled by the current developed inside the
capacitor as a result of the applied voltage excitation will lead
to similar results; i.e. observation of pinched-loops. This dual circuit
is depicted in Fig. \ref{fig:The-dual-of} where a current sensing
mechanism can be used to sense the current in the capacitor and re-use
it to control the parallel resistor. In solid-state devices, this
may naturally occur through leakage. In fact, the work of \cite{N2}
has already shown that in a $TiO_{2}$ memristor device, there must
exist two capacitors ($C_{ON}$ and $C_{OFF}$) parallel to the switch
on and switch off resistances ($R_{ON}$ and $R_{OFF}$) (see Fig.
5 of \cite{N2}) leading to the conclusion of that work that\emph{
``memory resistance and memory capacitance'' must co-exist.} From
a circuit theoretical point of view, it is known that in order to
measure a voltage it must be held on a capacitor for the duration
of the measurement. If this voltage is across a resistor, then there
must exist a physical or a parasitic capacitor to hold this voltage.
Therefore, it is literally impossible to ignore the capacitive effect
particularly in nano-scale devices \cite{N3}. In fact, we beg to
argue that the pinched-loop observed in nano-devices is either via
an inductive effect (following for example the topology of Fig. 1
above \cite{N1}) or via a capacitive effect (following the topology
of the dual circuit in Fig. \ref{fig:The-dual-of}, for example \cite{N2}).
The talk about a capacitance-free and/or inductance-free mem(resistor)
is fundamentally in-correct and obstructs the understanding of the
true mechanisms by which a pinched-loop can be created \cite{maundy2019correlation}
\cite{N88}. 

\section{Conclusion}

We investigated a new concept for generating pinched loops using the
current excitation of an inductive circuit. This work is a further
contribution to the body of research that confirms that pinched hysteresis
loops \textbf{are not unique} to memristors \cite{chua2014if} and
are not finger-prints of these devices as unfortunately widely advocated.
In addition, the question of whether the memristor is or is not a
fundamental element has been addressed by several authors (see for
example \cite{N4,N5}). In practice, the link between magnetic flux
and electric charge implies that one of them is the cause and the
other is the measurable effect. If the electric charge (magnetic flux)
is the cause, then there must exist a capacitive (inductive) effect
by which this charge (flux) is transported. Whether a pinched-loop
is observed or not is absolutely governed by satisfying the theory
of Lissajous figures \cite{N6,N7} between the current and voltage
signals.

\bigskip{}

\bibliographystyle{ieeetr}

\end{document}